\documentclass[conference]{IEEEtran}
\usepackage{amsmath,amsthm,amsfonts,amssymb}
\usepackage{graphicx}
\usepackage{braket} 
\usepackage{fancyhdr} 
\usepackage{soul} 
\usepackage{color} 
\usepackage[colorlinks,linkcolor=blue,anchorcolor=blue,citecolor=blue,urlcolor=black]{hyperref}
\usepackage{lipsum} 
\usepackage{cite} 

\newcommand{\bigchi}{\makebox{\large\ensuremath{\chi}}}
\DeclareMathAlphabet{\mathcal}{OMS}{cmsy}{m}{n}
\begin{document}
\title{\huge{Teleportation of Hybrid Entangled States with\\ Continuous-Variable Entanglement}}
\author{
	\IEEEauthorblockN{Mingjian He and Robert Malaney}\\
	\IEEEauthorblockA{School of Electrical Engineering  \& Telecommunications,\\
		The University of New South Wales,
		Sydney, NSW 2052, Australia.}}
\maketitle
\thispagestyle{fancy}

\renewcommand{\headrulewidth}{0pt}

\begin{abstract}
Hybrid entanglement between discrete-variable (DV) and continuous-variable (CV) quantum systems is an essential resource for heterogeneous quantum networks.
Our previous work showed that in lossy channels the teleportation of DV qubits, via CV-entangled states, can be significantly improved by a new protocol defined by a modified Bell state measurement at the sender.
This work explores whether a new, similarly modified, CV-based teleportation protocol can lead to improvement in the transfer of hybrid entangled states. To set the scene,  we first determine the performance of such a modified protocol in teleporting CV-only qubits, showing that significant improvement over traditional CV-based teleportation is obtained. We then explore similar modifications in the teleportation of a specific hybrid entangled state showing that significant improvement over traditional CV-based teleportation is again found. 
For a given channel loss, we find teleporting the DV qubit of the hybrid entangled state can always achieve higher fidelity than teleporting the CV qubit.
We then explore the use of various non-Gaussian operations in our modified teleportation protocol, finding  that, at a cost of lower success probability, quantum scissors provides the most improvement in the loss tolerance. Our new results emphasize that in lossy conditions, the quantum measurements undertaken at the sender can have a surprising and  dramatic impact on  CV-based  teleportation.
\end{abstract}

\maketitle

\thispagestyle{fancy}
\pagestyle{fancy}
\renewcommand{\headrulewidth}{0pt}

\section{Introduction}
Quantum information encoded in quantum states offers the premise of  information tasks that cannot be implemented by classical techniques. 
In quantum communications, the transmission of such information can be realized with photons. 
Within quantum optics, two different encoding schemes have been fostered due to the wave-particle duality of light. The discrete-variable (DV) encoding employs quantum states with finite dimensions, such as photon number and light polarization. DV schemes normally can accommodate photon loss with low system complexity. The continuous-variable (CV) encoding employs states with infinite Hilbert space, such as the quadrature fields of light. CV schemes normally benefit from  deterministic state preparation and the availability of ``off-the-shelf'' technologies based on efficient homodyne detectors.

Hybrid entanglement between DV and CV encoded states plays an important role in technologies that harness the benefits of both encoding schemes. 
An example of a hybrid entangled state is that created by the entanglement a CV Schrödinger-cat state and a DV photon-number state. Such a hybrid state enables the information conversion between quantum processors built upon different encodings. It finds applications in various quantum information tasks, such as teleportation \cite{park2012quantum,lee2013near,ulanov2017quantum,sychev2018entanglement}, remote state preparation between hybrid systems \cite{le2018remote,wen2021hybrid}, quantum steering \cite{cavailles2018demonstration}, and fault-tolerant quantum computing \cite{omkar2020resource}. 

Quantum state distribution is a prerequisite for many quantum information tasks. At times, direct transmission is impaired and teleportation-based distribution of the state offers better outcomes.
There are teleportation protocols that use different forms of entanglement between the sender and receiver as a means to distribute quantum states, e.g., DV entanglement \cite{bennett1993teleporting,marshall2014high}, CV entanglement \cite{vaidman1994teleportation,braunstein1998teleportation,milburn1999quantum,takeda2013gain}, or DV-CV hybrid entanglement \cite{brask2010hybrid,park2012quantum,lee2013near,lim2016loss,ulanov2017quantum,podoshvedov2019efficient,bich2022teleporting}.
The generation of CV entanglement in the form of two-mode squeezed vacuum (TMSV) states is well-established from both theoretical and experimental perspectives \cite{adesso2007entanglement}, making teleportation through TMSV states more feasible than other  entangled states.
However, in teleporting either qubit of a hybrid entangled state, \cite{do2021satellite} shows that for the loss region where teleportation  is better than direct transmission, teleportation via TMSV states cannot achieve high fidelity.

Our previous work \cite{he2022teleportation} shows that a modified  Bell state measurement (BSM) can improve the teleportation of DV qubits via TMSV  states. 
However, the feasibility of the modified measurement in teleporting CV qubits remains unknown.
As part of this  work, we close this gap. Then,
building upon our previous results \cite{he2022teleportation} on the teleportation of DV qubits, and our new results on the teleportation of CV qubits presented here, we move on to  our main focus---the teleportation of a hybrid entangled state.

Other  mechanisms for improving CV-based  teleportation (i.e., utilizing TMSV states)  include non-Gaussian operations on the modes of the TMSV state shared by the sender and receiver. Such operations are known to provide improved fidelity in the teleportation of both  DV states \cite{he2022teleportation} and CV states \cite{cochrane2002teleportation,dell2007continuous,yang2009entanglement,dellanno2010realistic,wang2015continuous-variable,xu2015enhancing,hu2017continuous-variable,villasenor2021enhancing,asavanant2021wave, kumar2022experimental}.
We  extend our study of entangled state teleportation using modified measurements at the sender, to include the use of various non-Gaussian operations (photon subtraction, photon addition, photon catalysis, and quantum scissors) at both the sender and the receiver.
Beyond the independent application of non-Gaussian operations on both modes of the TMSV state, we  also investigate  delocalized forms of non-Gaussian operations, which have been shown to significantly enhance the entanglement of certain entangled states \cite{biagi2020entangling, lan2021two,dat2022entanglement,liu2022optimal}.

To summarize, the main contributions of this work are as follows:
(i) We compare traditional CV-based teleportation  with a new CV-based teleportation protocol built on a modified BSM, in the context of the teleportation of CV qubits. We determine the conditions under which our protocol provides better performance in terms of fidelity.
(ii) We extend this comparison into the context of  CV-based teleportation of a hybrid entangled state, investigating the separate teleportation of the DV and the CV qubits of the hybrid state.
(iii) Building on our new protocol, we then compare the use of  different non-Gaussian operations, as applied to the modes of the TMSV states utilised, determining which operation provides the most improved fidelity under various conditions.

The remainder of this paper is organized as follows. In Section~\ref{sec:2} we present the teleportation protocols and study their performance in the teleportation of CV qubits.
In Section~\ref{sec:3} we build on the above,  investigating the teleportation of a hybrid entangled state.
In Section~\ref{sec:4} we introduce additional non-Gaussian operations and study the use of such operations in the teleportation of the entangled state.
Section~\ref{sec:5} concludes our work.
 
\section{Teleportation of CV qubits}\label{sec:2}

In the generic  teleportation scheme between a sender (A) and receiver (B) discussed in this work we define the following entities.
 The \emph{input mode} at A  is the quantum state (or mode of a two-mode state) that is to be teleported; the \emph{resource state} is the two-mode state shared by  A and B  used for invoking the teleportation,  and the \emph{output mode} is the quantum state (or mode of a two-mode state)  that is the output of the teleportation (in the ideal case, identical to the input mode). The \emph{channel} refers to the link used to transmit a mode of the resource state to  A or B.
To set the scene for the teleportation of a mode of a hybrid entangled state, let us first discuss a  simpler scenario---the teleportation of a CV qubit.
Consider the CV qubit
\begin{equation}\label{eq:catqbket}
	\ket{\mathrm{in}}=\cos\frac{\theta}{2}\ket{\mathrm{cat_-}}+\exp(i\phi)\sin\frac{\theta}{2}\ket{\mathrm{cat_+}},
\end{equation}
where $0\leq\theta\leq\pi$, $0\leq\phi\leq2\pi$, and $\ket{\mathrm{cat_\pm}}$ are the Schrödinger-cat states given by
\begin{equation}\label{eq:cat}
	\ket{\mathrm{cat_\pm}}=(\ket{\alpha}\pm\ket{-\alpha})/\sqrt{N_\pm},
\end{equation}
where $\ket{\alpha}$ and $\ket{-\alpha}$ are coherent states with complex amplitudes $\pm\alpha$, and 
\begin{equation}
N_\pm=2[1\pm\exp(-2|\alpha|^2)],
\end{equation}
are normalization constants.
Without loss of generality, we will assume $\alpha$ is real. 


 Now, consider the original CV teleportation protocol  \cite{vaidman1994teleportation} applied to the teleportation of the above CV qubit.
In this protocol, a TMSV state is used as the resource state.
We label the two modes of the resource state by 1 and 2 and the input mode by 3.
A CV Bell state measurement (CV-BSM) is performed on modes 1 and 3. In such a measurement,
mode 1 is coupled with mode 3 at a 50:50 beam splitter.
The $p$-quadrature of one output mode of the beam splitter and the $q$-quadrature of the other output mode are measured by two homodyne detectors.
Based on the measurement result, a displacement operation (with displacement gain factor $g$) is applied to mode 2.
In the absence of channel loss, with infinite initial squeezing mode 2 approaches the input mode.
Henceforth, we refer to this CV-BSM based teleportation protocol as the \emph{CV-BSM protocol}.

\begin{figure}
	\centering
	\includegraphics[width=0.98\linewidth]{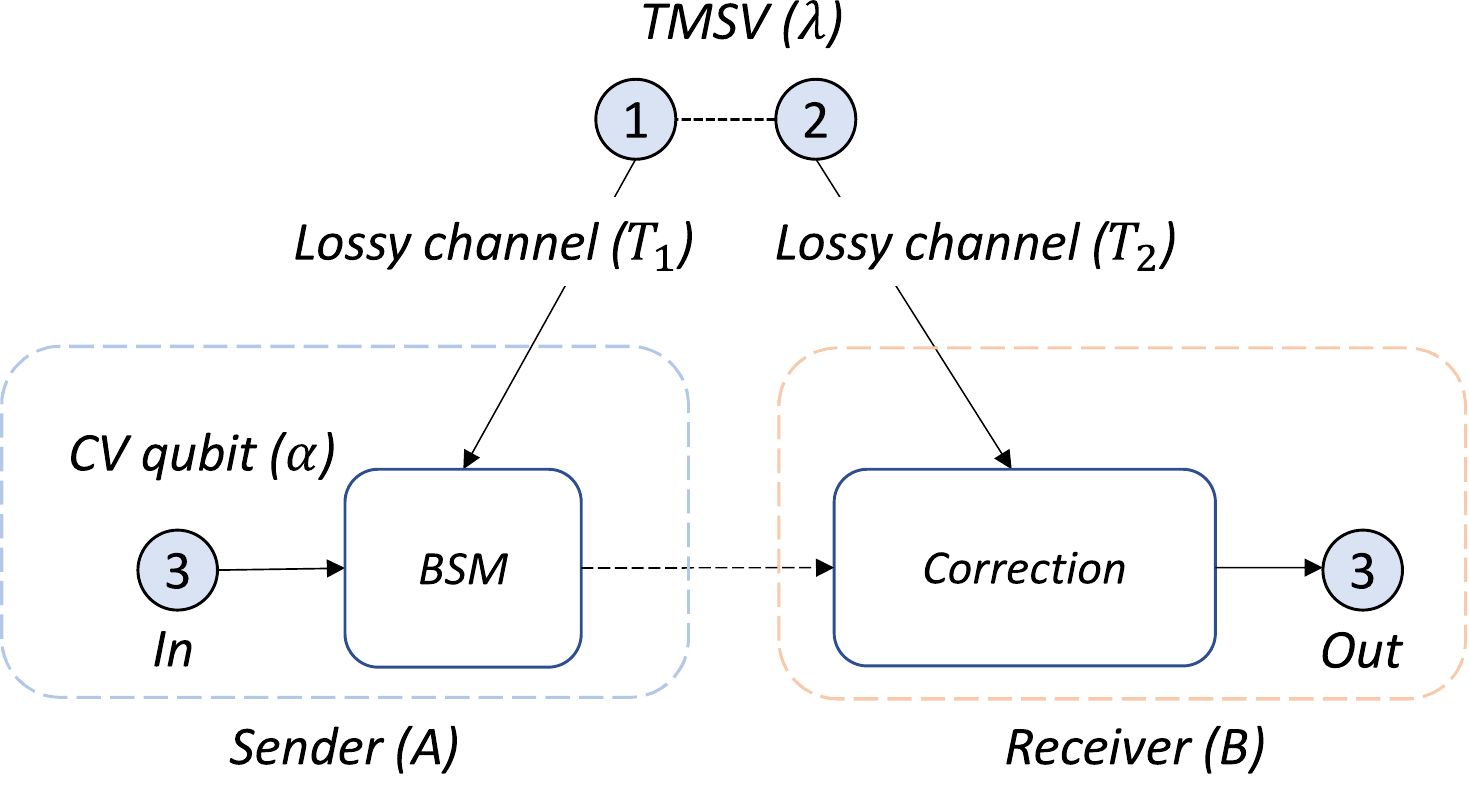}
	\caption{The teleportation of a CV qubit from the sender A to the receiver B. An initial TMSV state with modes 1 and 2 is first prepared at a middle station and then transmitted to A and B via two independent lossy channels. This evolved TMSV state is then used as the resource state. In teleportation, a BSM is applied to modes 1 and 3 at the sender and a correction operation is applied to mode 2 at the receiver.}
	\label{fig:diagprotocols}
\end{figure}

The relation between our input mode and  output mode can be described using the characteristic function formalism.
As shown in Fig.~\ref{fig:diagprotocols},
consider a realistic scenario where the sender and receiver do not share any resource state before teleportation, and an initial TMSV state is first prepared at a middle station. 
The characteristic function for the initial TMSV state can be written as
\begin{equation}
	\begin{aligned}
		\bigchi_{\mathrm{TMSV}}(\xi_1, \xi_2)=&\exp \Big\{ -\frac{1}{2}\Big[\frac{1+\lambda^2}{1-\lambda^2} (|\xi_1|^2 + |\xi_2|^2 )\\ 
		&- \frac{2\lambda}{1-\lambda^2}(\xi_1 \xi_2 + \xi_1^* \xi_2^*) \Big] \Big\},	
	\end{aligned}
	\label{eq:CF_TMSV}
\end{equation}
where $\xi_1$ and $\xi_2$ are complex variables, $\lambda=\tanh{r}$, and $r>0$ is the squeezing parameter for the TMSV state.
The two modes of the TMSV state are then sent through two independent lossy channels characterized by the transmissivities $T_1$ and $T_2$.
The characteristic function for the TMSV state after the channel transmission can be written as \cite{li2011time}
\begin{equation}\label{eq:tmsv_lossy}
	\begin{aligned}
		\bigchi_{\mathrm{TMSV}}'(\xi_1, \xi_2)=&\exp{\left\{-\frac{1}{2}\left[(1 - T_1)|\xi_1|^2+(1 - T_2)|\xi_2|^2\right]\right\}}\\
		&\times\bigchi_{\mathrm{TMSV}}(\sqrt{T_1}\xi_1,\sqrt{T_2}\xi_2).
	\end{aligned}
\end{equation}
The distributed TMSV state with the above characteristic function is then used as the  resource state.

Let $\bigchi_{\mathrm{in}}(\xi)$ be the characteristic function of the input mode [with the state vector given by Eq.~(\ref{eq:catqbket})], which can be calculated from the definition $\bigchi_{\mathrm{in}}(\xi)=\operatorname{tr}\{\ket{\mathrm{in}}\bra{\mathrm{in}}\hat{D}(\xi)\}$, where $\hat{D}(\xi)$ is the displacement operator of mode 3.
The characteristic function of the averaged output mode can then be written as \cite{dell2010realistic}
\begin{equation}\label{eq:cfinandout}
	\bigchi_{\mathrm{out}}(\xi)= 
	\bigchi_{\mathrm{in}}(g \xi) \bigchi_{\mathrm{TMSV}}'(g  \xi^*, \xi),
\end{equation}
where the gain satisfies $g>0$,  and the average has been taken over the measurement outcome of the CV-BSM.
Let $\hat{\rho}_\mathrm{in}=\ket{\mathrm{in}}\bra{\mathrm{in}}$ and $\hat{\rho}_\mathrm{out}$ be the density operators for the input and output modes, respectively.
The teleportation fidelity, which measures the closeness between $\hat{\rho}_\mathrm{in}$ and $\hat{\rho}_\mathrm{out}$, is then defined as \cite{jozsa1994fidelity}
\begin{equation}\label{eq:fidelityoriginal}
	\mathcal{F}=\left(\mathrm{tr}\left\{\sqrt{\sqrt{\hat{\rho}_\mathrm{out}}\hat{\rho}_\mathrm{in}\sqrt{\hat{\rho}_\mathrm{out}}}\right\}\right)^2.
\end{equation}
Since the input mode is pure, for the CV-BSM protocol, the teleportation fidelity defined above can be re-written as \cite{chizhov2002continuous}
\begin{equation}\label{eq:fidelitycf}
	\bar{\mathcal{F}} = \frac{1}{\pi} \int d^2 \xi
	\bigchi_\mathrm{in}(\xi)
	\bigchi_\mathrm{out}(-\xi),
\end{equation}
where the averaging symbol $\bar\cdot$ is adopted here since $\bar{\mathcal{F}}$ can be viewed as the fidelity averaged over the probability distribution of the outcome of the CV-BSM.

Next, consider a modified teleportation protocol, which uses the same resource state as before, but with a different strategy on the measurement of the input mode and the correction of the output mode.
In the Fock basis, the initial TMSV state prepared at the middle station can be written as
\begin{equation}
	\ket{\mathrm{TMSV}}=\sqrt{1-\lambda^2}\sum_{n=0}^{\infty}\lambda^n\ket{nn}_{12}.
\end{equation}
The initial TMSV state is then sent to the sender and receiver. The two independent lossy channels alter the TMSV state to \cite{sabapathy2011robustness}
\begin{equation}\label{eq:rhotmsvlossy}
	\hat{\rho}_{\mathrm{TMSV}}'=\mathcal{T}_1\circ\mathcal{T}_2(\hat{\rho}_{\mathrm{TMSV}}),
\end{equation}
where $\hat{\rho}_{\mathrm{TMSV}}=\ket{\mathrm{TMSV}}\bra{\mathrm{TMSV}}$, $\circ$ represents the composition of transformations, and the transformation $\mathcal{T}_k$ ($k=1, 2$) is defined as
\begin{equation}\label{eq:lossyopeator}
	\mathcal{T}_k(\hat\rho):=
	\sum_{l=0}^\infty 
	\hat{G}^{(l)}_{k}
	\hat{\rho}
	\hat{G}^{(l)\dagger}_{k},
\end{equation}
with a set of Kraus operators given by
\begin{equation}\label{eq:channelkraus}
	\hat{G}_{k}^{(l)}=\frac{\hat{a}^l_k\sqrt{T_k}^{\hat{a}_k^\dagger\hat{a}_k}}{\sqrt{l!}}\sqrt{\frac{1-T_k}{T_k}}^l,
\end{equation}
and $\hat{a}_k$ the annihilation operator of mode $k$. 

A hybrid BSM (H-BSM) is then performed on modes 1 and 3. 
This H-BSM can be represented by a complete set of projectors, $\mathcal{P}=\{\hat{P}_1,\hat{P}_2,\hat{P}_3,\hat{P}_4,\hat{P}_\mathrm{fail}\}$, where
\begin{equation}\label{eq:comhbsm}
\begin{aligned}
    &\hat{P}_1=\ket{\Phi^+}\bra{\Phi^+},\ 
    \hat{P}_2=\ket{\Phi^-}\bra{\Phi^-},\\
    &\hat{P}_3=\ket{\Psi^+}\bra{\Psi^+},\ 
    \hat{P}_4=\ket{\Psi^-}\bra{\Psi^-},\\
	&\hat{P}_\mathrm{fail}=I-\hat{P}_1-\hat{P}_2-\hat{P}_3-\hat{P}_4,
\end{aligned}
\end{equation}
$I$ is the identity matrix, and
\begin{equation}
	\begin{aligned}
		\ket{\Phi^\pm}&=\frac{1}{\sqrt2}\left(\ket{00}\pm\ket{11}\right),\\
		\ket{\Psi^\pm}&=\frac{1}{\sqrt2}\left(\ket{01}\pm\ket{10}\right).
	\end{aligned}
\end{equation}
Modes 1 and 3 are projected into one of the four Bell states for $\hat{P}_1$ to $\hat{P}_4$. The success probability for the H-BSM, which is the probability that the modes are projected into one Bell state, is then defined as
\begin{equation}
    P_\mathrm{BSM}=\sum_{i=1}^4 \operatorname{tr}\{\hat{P}_i \hat{\rho}_{\mathrm{in}}\otimes \rho'_{\mathrm{TMSV}}\}.
\end{equation}
We refer to the H-BSM based teleportation protocol as the \emph{H-BSM protocol}.

In the absence of channel loss, after a truncation (to $\ket{0}$ and $\ket{1}$) is applied on either of its modes, an infinitely squeezed TMSV state approaches the Bell state $\ket{\Phi^+}$. 
This is at the core of the H-BSM protocol.
A truncation that projects a high dimensional state into a lower dimension is non-deterministic.
Consequently, the H-BSM is non-deterministic and its success probability decreases with increasing mean photon numbers in either of modes 1 and 3.
We define the total success probability for the H-BSM protocol as
\begin{equation}
	P_{\mathrm{total}}=P_{\mathrm{BSM}}P_{\mathrm{operation}},
\end{equation}
where $P_{\mathrm{operation}}$ is the success probability for additional operations.
Mode $2$ will be discarded when the protocol fails.
For the H-BSM protocol considered in this section $P_{\mathrm{total}} = P_{\mathrm{BSM}}$ because no additional operation is adopted.

Note, there is a degenerate form of the H-BSM, which can be represented by a different set of projectors \cite{lombardi2002teleportation}, $\mathcal{P}'=\{\hat{P}_3,\hat{P}_4,I-\hat{P}_3-\hat{P}_4\}$. We refer to this form of H-BSM, which can only project the modes into $\ket{\Psi^\pm}$, as the incomplete H-BSM. Unless specified otherwise, we will assume the complete H-BSM is adopted.
Possible ways to  implement the complete H-BSM can be found in \cite{grice2011arbitrarily, he2022teleportation}.
Depending on the result of the H-BSM, a transformation that is unitary in the space spanned by $\ket{0}$ and $\ket{1}$ 
is applied to mode $2$ of the  resource state. 
In this way, mode $2$ is converted to a state that approximates the input mode.


For the H-BSM protocol, since the input mode is pure, the teleportation fidelity defined by Eq.~(\ref{eq:fidelityoriginal}) reduces to \cite{chizhov2002continuous}
\begin{equation}\label{eq:fidelityden}
	\mathcal{F} = \mathrm{tr}\left\{\hat{\rho}_\mathrm{in}\hat{\rho}_\mathrm{out}\right\},
\end{equation}
which is equivalent to the fidelity given by Eq.~(\ref{eq:fidelitycf}). 
The fidelity averaged over the four possible outcomes of the H-BSM (and normalized by $P_{\mathrm{total}}$), $\bar{\mathcal{F}}$, will be used as the performance metric for the protocol.

\begin{figure}
	\centering
	\includegraphics[width=0.9\linewidth]{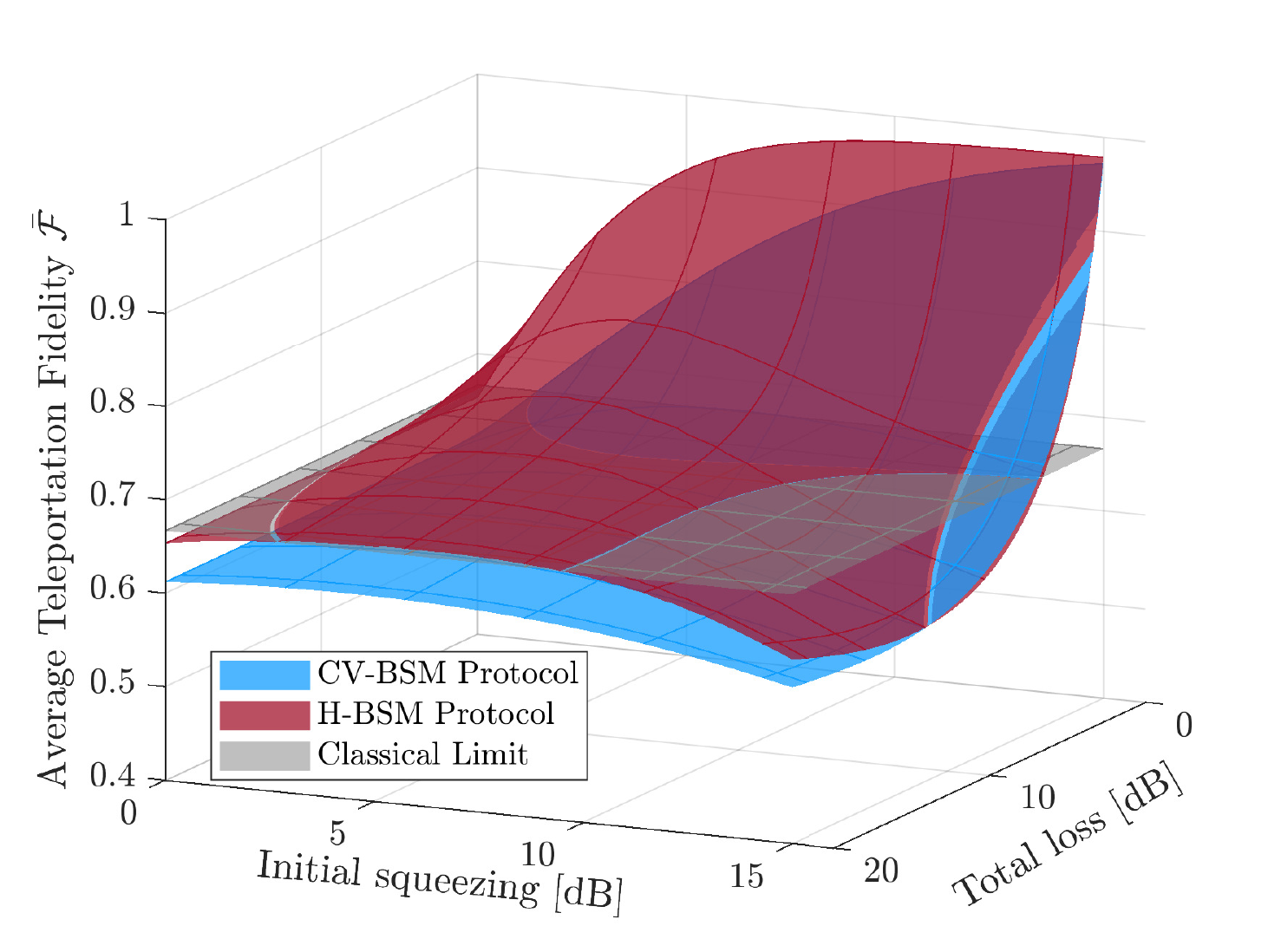}\\
	(a) $\alpha=0.5$\\
	\includegraphics[width=0.9\linewidth]{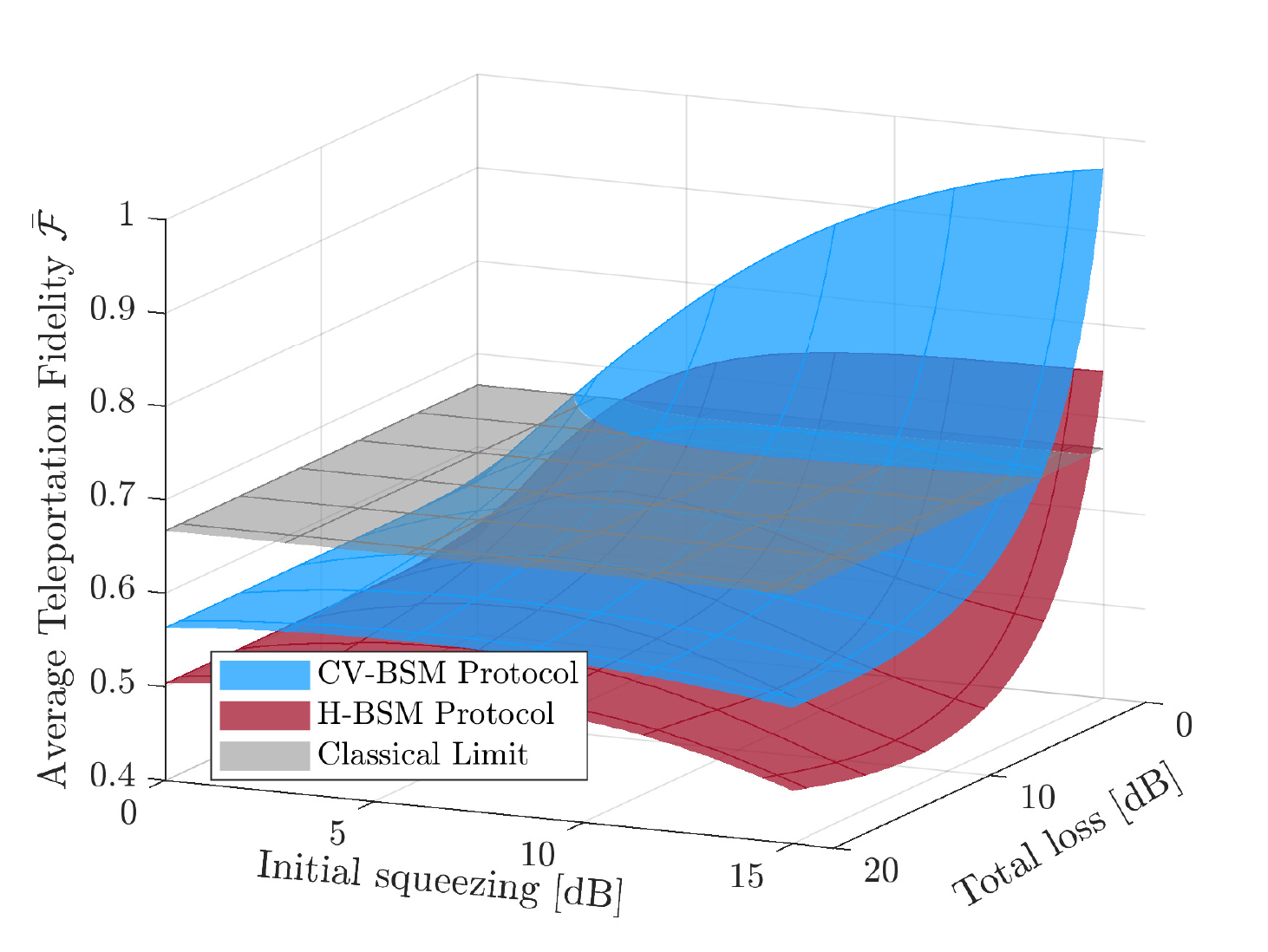}\\
	(b) $\alpha=1$
	\caption{The average fidelity for the resultant CV qubit, which is teleported using the CV-BSM protocol or the H-BSM protocol. For the CV-BSM protocol (the blue surface), the displacement gain $g$ is set so as to maximize $\bar{\mathcal{F}}$. The gray surface represents the $2/3$ classical limit on the teleportation of qubit. The opacity of all surfaces are set to $75\%$.}
	\label{fig:figbella05}
\end{figure}

Consider the scenario where the lossy channels are symmetric, i.e., $T_1=T_2=T$. In the following calculations,
the channel loss in decibel units is determined via $T\mathrm{[dB]}=-10\log_{10}\left(T\right)$.
The total channel loss of the two channels is then $2T\mathrm{[dB]}$.
The initial squeezing of the TMSV states in decibel units is determined via $r\mathrm{[dB]}=-10\log_{10}[\exp(-2r)]$.
In all calculations, the initial TMSV state (and the input mode) is first truncated to a finite dimension such that $\mathrm{tr}(\hat{\rho}_{\mathrm{TMSV}})>0.95$ is satisfied. The truncated state is then normalized.

We assume the CV qubits to be teleported are evenly distributed on the Bloch sphere: the probability distribution function for the parameters $\theta$ and $\phi$ is then $P(\theta,\phi)=(\sin\theta)/(4\pi)$. The average fidelity (averaged over the probability distribution of the outcome of the CV-BSM or H-BSM and $P(\theta, \phi)$) for the CV-BSM protocol and the H-BSM protocol are shown in Fig.~\ref{fig:figbella05}.
The $2/3$ classical limit on the teleportation of qubits is also shown as a benchmark for the protocols.
From Fig.~\ref{fig:figbella05}a we can see the H-BSM protocol provides fidelity higher than the CV-BSM protocol when $\alpha$ is small. 
In terms of surpassing the classical limit, the H-BSM protocol can tolerate higher channel loss than the CV-BSM protocol.
As can be seen from Fig.~\ref{fig:figbella05}b, for large $\alpha$, the CV-BSM protocol provides fidelity higher than the H-BSM protocol over the entire parameter space.
For the H-BSM protocol, the achievable fidelity is far from unity even with high initial squeezing because the protocol can only retrieve the information in the $\ket{0}$ and $\ket{1}$ components of the input mode.
Both protocols fail to beat the classical limit when the channel loss exceeds a small threshold. Although not shown,
Similar conclusions hold for the scenario of asymmetric channel losses.

\section{Teleportation of hybrid entangled states}\label{sec:3}
We now move on to main topic of this work---the teleportation of hybrid entangled states. Consider the hybrid state  
 \cite{park2012quantum}
\begin{equation}\label{eq:entket}
	\ket{\mathrm{in}}=\frac{1}{\sqrt{2}}\left(\ket{\mathrm{cat}_-}\ket{0}+\ket{\mathrm{cat}_+}\ket{1}\right),
\end{equation}
where $\ket{\mathrm{cat}_\pm}$ are the cat states defined by Eq.~(\ref{eq:cat}).
We label the CV qubit and the DV qubit of the hybrid entangled state by modes 3 and 4, respectively. The resource state will again be labeled by modes 1 and 2.

For both the CV-BSM protocol and the H-BSM protocol, the procedures of teleportation of one qubit of a hybrid entangled state is similar to that of the CV qubit discussed in Section~\ref{sec:2}. However some difference exist, which we now 
 briefly discuss. 
Consider the teleportation of the DV qubit of the hybrid entangled state (i.e., mode 4 as the input mode), for the CV-BSM protocol, a CV-BSM is performed on modes 1 and 4. A displacement operation is applied to mode 2 of the resource state.
Let $\bigchi_{\mathrm{in}}(\xi_3, \xi_4)$ be the characteristic function of the hybrid entangled state given by Eq.~(\ref{eq:entket}), which can be calculated using the definition $\bigchi_{\mathrm{in}}(\xi_3, \xi_4)=\operatorname{tr}\{\ket{\mathrm{in}}\bra{\mathrm{in}}\hat{D}(\xi_3)\hat{D}(\xi_4)\}$, where $\hat{D}(\xi_3)$ and $\hat{D}(\xi_4)$ are the displacement operators of the two modes of the hybrid entangled state.
The characteristic function of the averaged output of teleportation can then be written as \cite{dell2019non} 
\begin{equation}\label{eq:cfinandouteng}
	\bigchi_{\mathrm{out}}(\xi_3, \xi_4)= 
	\bigchi_{\mathrm{in}}(\xi_3, g \xi_4) \bigchi_{\mathrm{TMSV}}'(g  \xi_4^*, \xi_4),
\end{equation}
where again $g$ is the displacement gain factor.
The fidelity between the initial hybrid entangled state (with modes 3 and 4) and the state after teleportation (with modes 2 and 3) can be written as
\begin{equation}\label{eq:fidelitycfeng}
	\bar{\mathcal{F}} = \frac{1}{\pi^2} \int d^2 \xi_3 d^2 \xi_4
	\bigchi_\mathrm{in}(\xi_3, \xi_4)
	\bigchi_\mathrm{out}(-\xi_3, -\xi_4).
\end{equation}
For the H-BSM protocol, an H-BSM is performed on modes 1 and 4. The fidelity can be calculated using Eq.~(\ref{eq:fidelityden}).

\begin{figure}
	\centering
	\includegraphics[width=.85\linewidth]{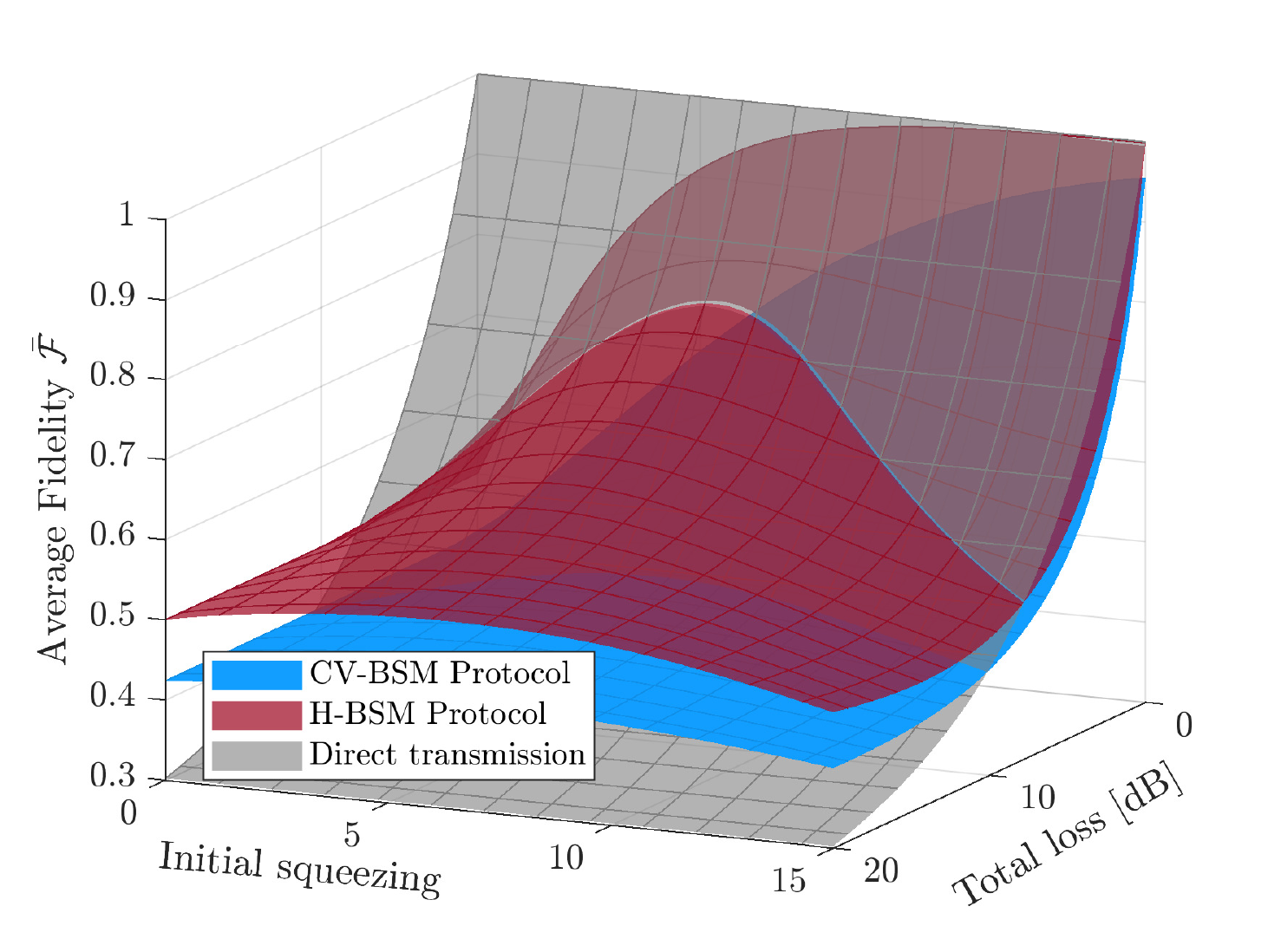}
	\caption{The average fidelity for the hybrid entangled state, of which the DV qubit is teleported using the CV-BSM protocol or the H-BSM protocol. 
	The results shown are independent of $\alpha$.
		For the CV-BSM protocol (the blue surface), the displacement gain $g$ are set so as to maximize $\bar{\mathcal{F}}$. 
		For comparison, the gray surface shows the fidelity for the directly  transmitted hybrid entangled state.
		For better illustration of the overlapped area of the three surfaces, the opacity of the red and gray surfaces is set to $75\%$.}
	\label{fig:figbellandhomoswappingdvtp}
\end{figure}

Again, consider the scenario where the lossy channels are symmetric.
Fig.~\ref{fig:figbellandhomoswappingdvtp} shows the average fidelity of the teleportation of the DV qubit of the hybrid entangled state.
For both the CV-BSM protocol and the H-BSM protocol, $\bar{\mathcal{F}}$ is independent of $\alpha$.
The fidelity for the hybrid entangled state with its DV qubit being directly transmitted over the two lossy channels (from the sender through the two lossy channels to the receiver) is also included as a benchmark.
For the CV-BSM protocol, the displacement gain is set so as to maximize $\bar{\mathcal{F}}$.
Similar to the results obtained in our previous work \cite{he2022teleportation}, the H-BSM protocol provides higher $\bar{\mathcal{F}}$ over the entire regions of parameter space considered.

\begin{figure}
	\centering
	\includegraphics[width=.9\linewidth]{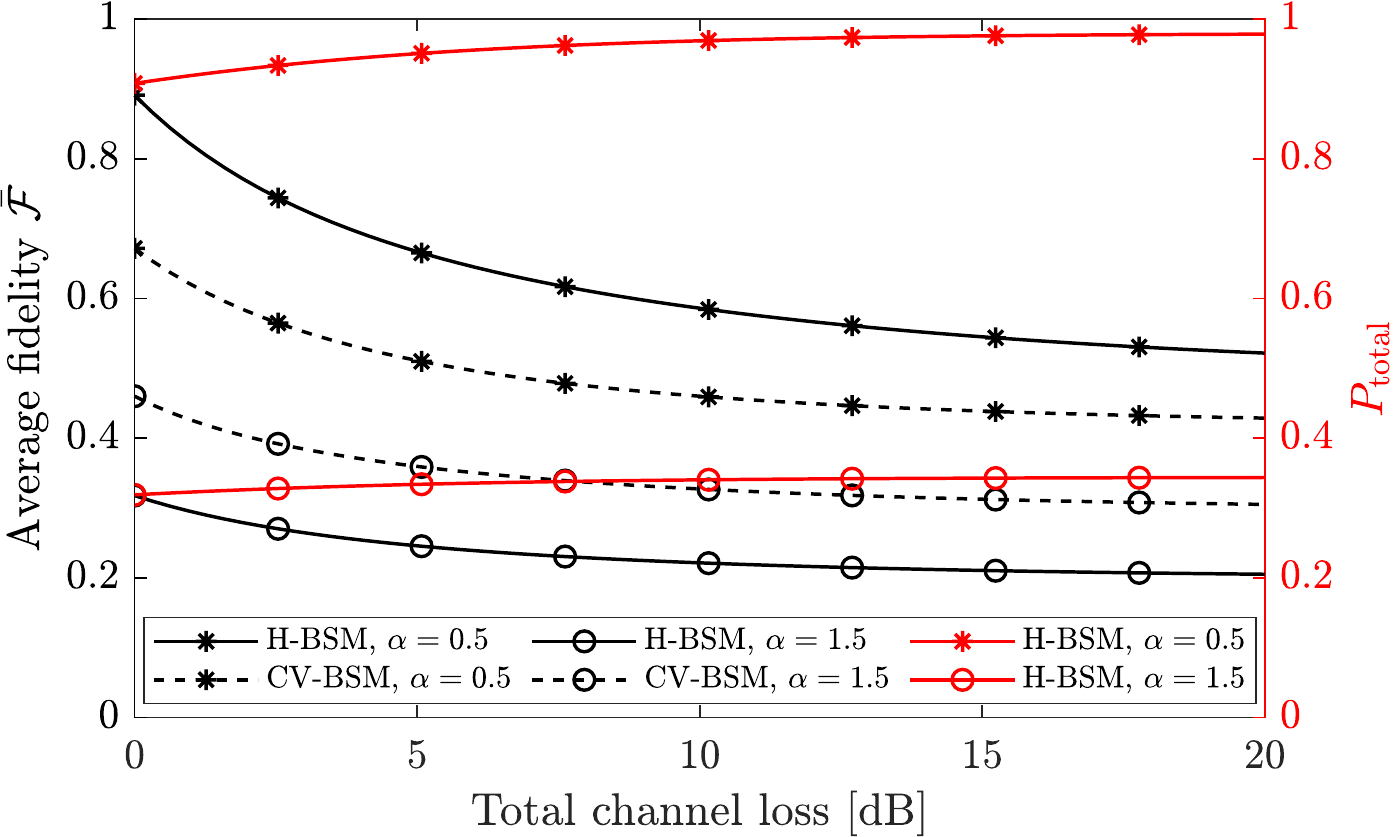}\\
	(a) Initial squeezing $=5$dB \\
	\includegraphics[width=.9\linewidth]{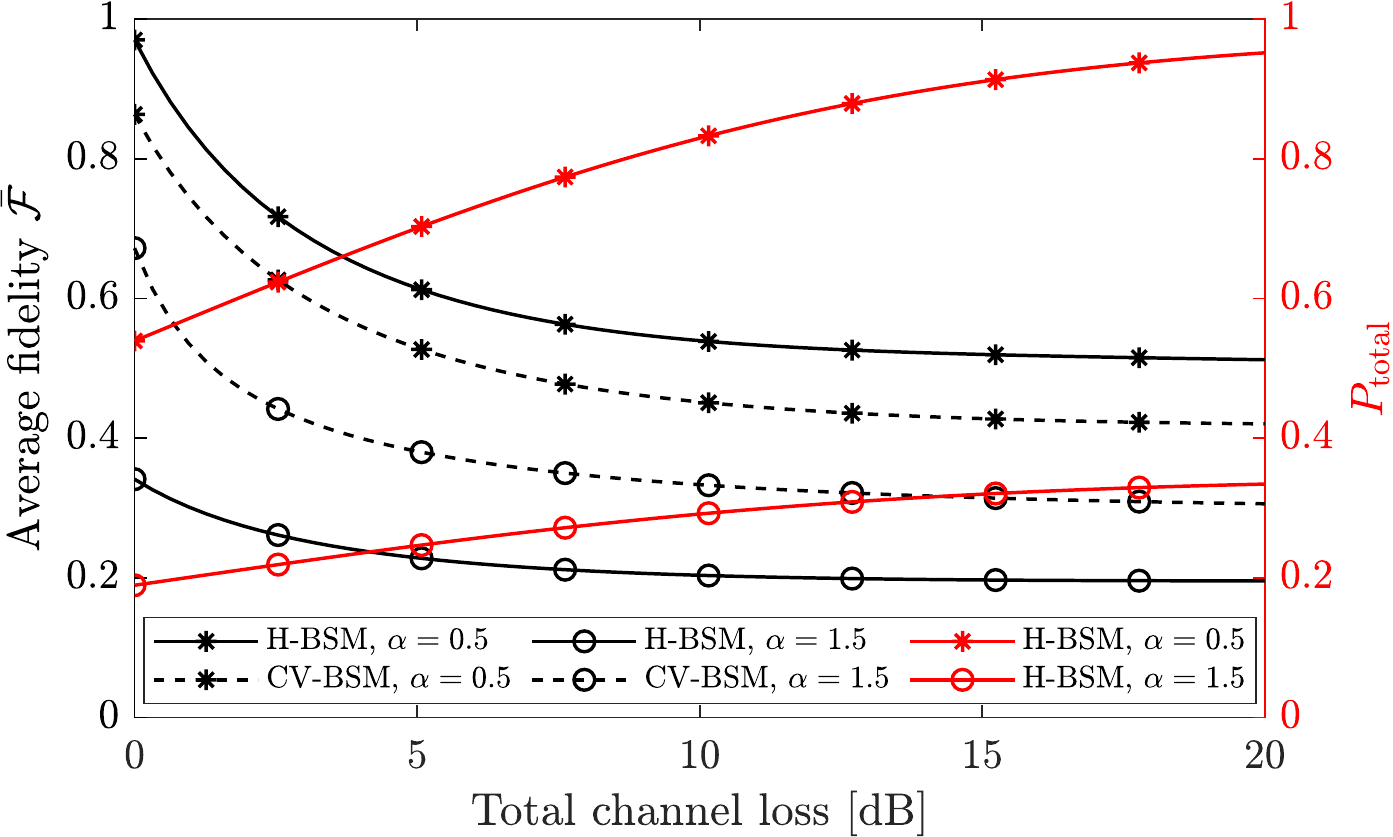}\\
	(b) Initial squeezing $=10$dB
	\caption{[(a), (b)] The black curves show the average fidelity for the resultant hybrid entangled state, of which the CV qubit is teleported using the CV-BSM protocol or the H-BSM protocol. For the CV-BSM protocol (the dashed curves), the displacement gain $g$ are set so as to optimize $\bar{\mathcal{F}}$. The red curves show the total success probability for the H-BSM protocol.}
	\label{fig:figbellandhomoswappingbsqz}
\end{figure}

Next, consider the teleportation of the CV qubit of the hybrid entangled state (i.e., mode 3 as the input mode).
In Fig.~\ref{fig:figbellandhomoswappingbsqz}, the black curves compare the average fidelity for the CV-BSM protocol and the H-BSM protocol with fixed $\alpha$ and $r$.
The H-BSM protocol can achieve higher fidelities than the CV-BSM protocol when $\alpha=0.5$ because the CV qubit of the hybrid entangled state approaches a DV qubit for small $\alpha$.
The total success probability $P_\mathrm{total}$ for the H-BSM protocol increases with increasing total channel loss due to the decrease of the mean photon number of the resource state.
When an incomplete H-BSM is adopted, $P_\mathrm{total}$ will be halved and $\bar{\mathcal{F}}$ will be slightly changed with a difference less than $5\%$.

\begin{figure}
	\centering
	\includegraphics[width=.85\linewidth]{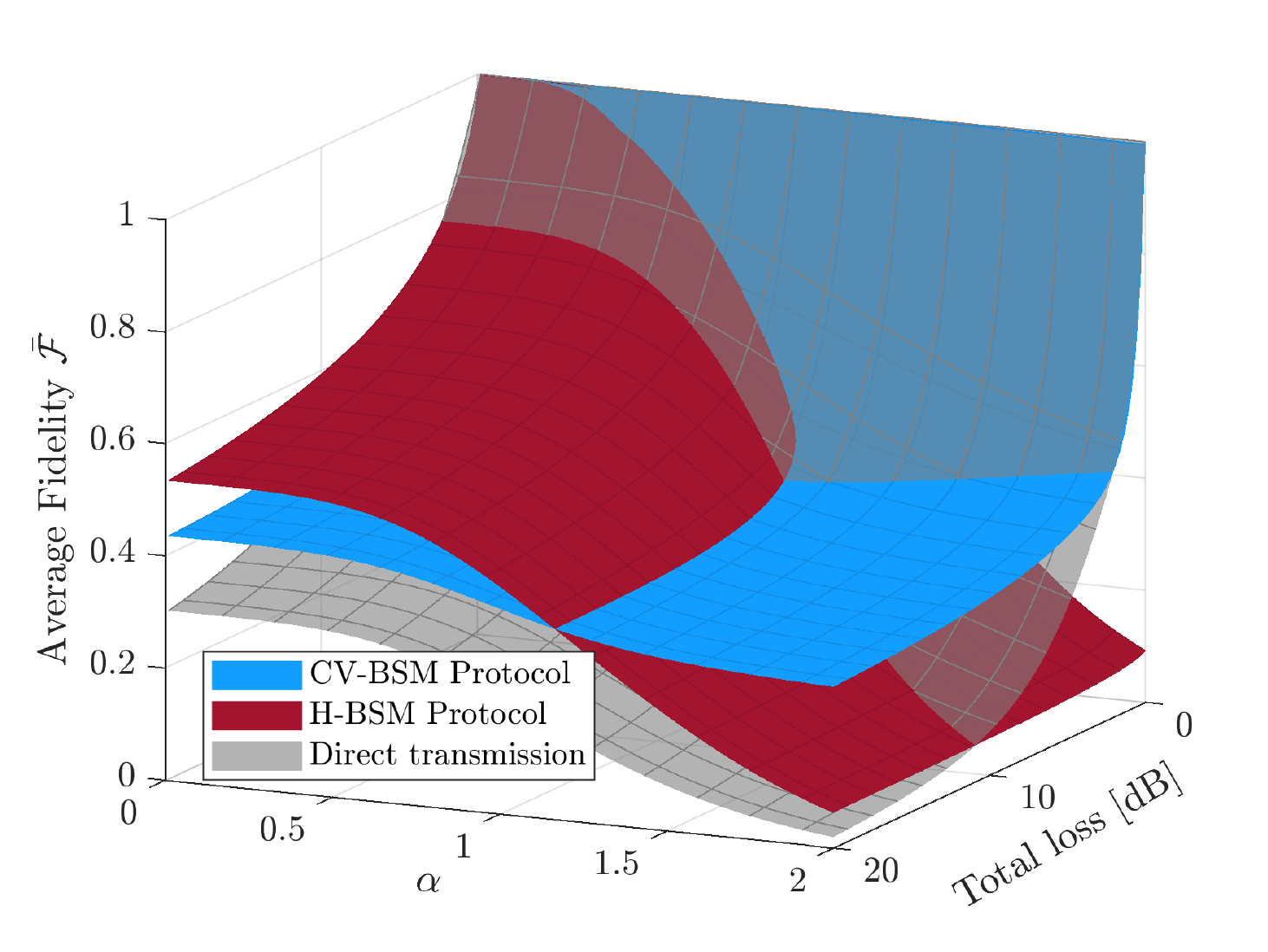}
	\caption{The maximized average fidelity for the resultant hybrid entangled state, of which the CV qubit is teleported using the CV-BSM protocol or the H-BSM protocol. 
	The fidelity for the hybrid entangled state with its CV qubit being directly transmitted over the two lossy channels  is also included as a benchmark.
	For the H-BSM protocol (the red surface), the initial squeezing of the TMSV state $r$ is set so as to maximize $\bar{\mathcal{F}}$.
	For the CV-BSM protocol (the blue surface), both $r$ and the displacement gain $g$ are set so as to maximize $\bar{\mathcal{F}}$. 
	An upper limit of 16dB is set for $r$[dB] in optimization.
	The shaded red and blue areas indicate that the direct transmission provides higher fidelity.}	
	\label{fig:figbellandhomoswappingopt}
\end{figure}

To better compare the two protocols, in Fig.~\ref{fig:figbellandhomoswappingopt}, for both protocols, for given total channel loss and $\alpha$ of the hybrid entangled state, the  squeezing of the initial TMSV state is adjusted independently so as to maximize the average fidelity. 
For the CV-BSM protocol, The displacement gain is also optimized.
The fidelity for the hybrid entangled state with its CV qubit being directly transmitted over the two lossy channels (from the sender through the two lossy channels to the receiver) is also included as a benchmark.
The H-BSM protocol is less sensitive to the channel loss but more sensitive to $\alpha$ in comparison with the CV-BSM protocol.
For the H-BSM protocol, a fidelity approaching unity is not achievable even with infinite squeezing when $\alpha>0$.
Independent of channel loss, the fidelity of the H-BSM protocol decreases significantly as $\alpha$ increases and drops below the  fidelity of the CV-BSM for large $\alpha$.
Both protocols fail to beat the direct transmission scheme when the total channel loss is small (e.g., $<5$dB when $\alpha=1$).

\section{Teleportation of Hybrid Entangled States with Non-Gaussian Resource States}\label{sec:4}

\begin{figure}
	\centering
	\includegraphics[width=.9\linewidth]{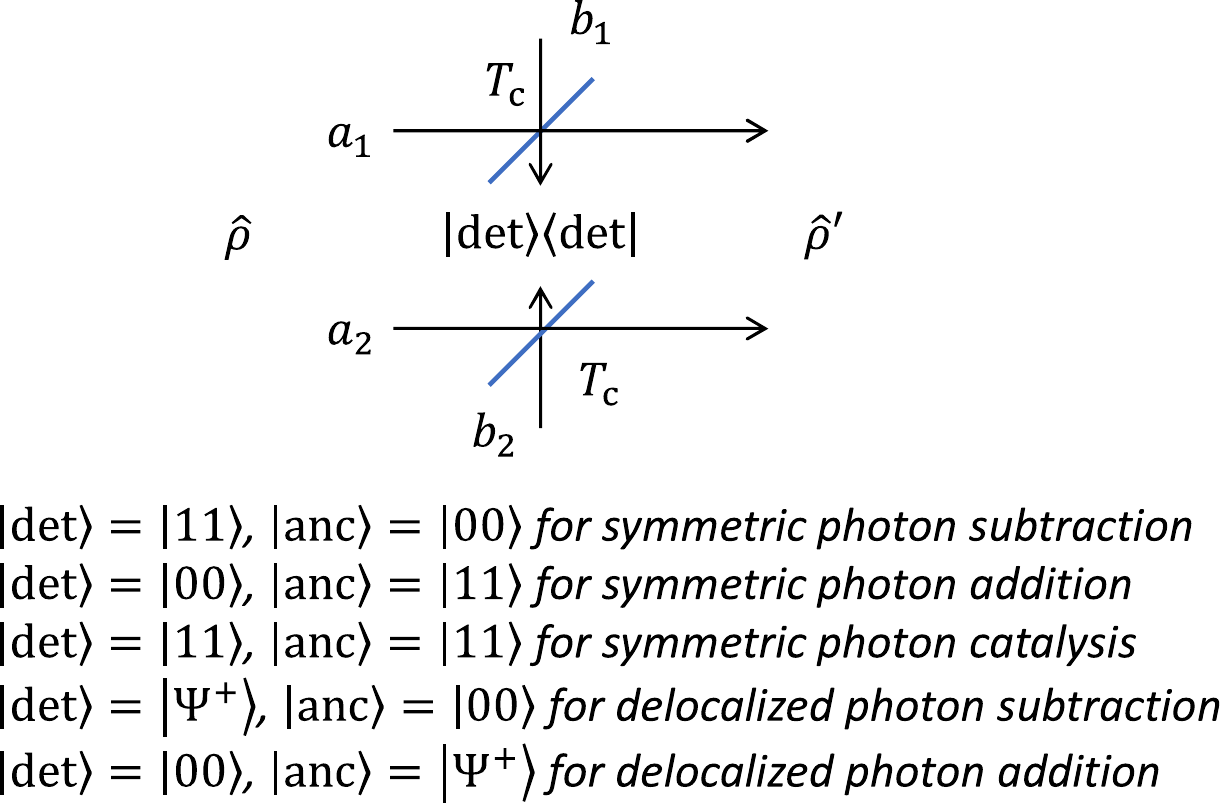}
	\caption{The symmetric/delocalized non-Gaussian operations applied to the two modes denoted by $a_1$ and $a_2$ of an input $\hat{\rho}$.
		The ancillary modes are denoted by $b_1$ and $b_2$ and in the state $\ket{\mathrm{anc}}$. 
		The state $\ket{\mathrm{anc}}=\ket{\Psi^+}=(\ket{01}+\ket{10})/\sqrt2$ can be prepared by mixing a mode in the state $\ket{1}$ and a mode in the state $\ket{0}$ at a 50:50 beam splitter. 
		The detection of the state $\ket{\mathrm{det}}=\ket{\Psi^\pm}$ can be realized by first mixing the modes to be detected at a 50:50 beam splitter.
		Two photon number detectors are in placed at the outputs of the beam splitter.
		The detection is successful if one photon has been detected at one output of the beam splitter.}
	\label{fig:diagdeng}
\end{figure}

Let us now consider the use of additional non-Gaussian operations in the H-BSM protocol.
In this section, we focus on the scenario where the CV qubit of the hybrid entangled state is teleported.
We assume the same type of non-Gaussian operations are performed on both modes of the TMSV state, before or after the channel transmission.
We refer to such operations as the symmetric operations.
For all non-Gaussian operations considered in this work, the resultant state after the operation can be written as
\begin{equation}
    \hat{\rho}'=\frac{1}{P_\mathrm{operation}}\hat{O}\hat{\rho}\hat{O}^\dagger,
\end{equation}
where $\hat{O}$ is the operator for the operation, and $P_\mathrm{operation}=\operatorname{tr}\{\hat{O}\hat{\rho}\hat{O}^\dagger\}$ is the corresponding success probability.

Consider two types of widely studied non-Gaussian operations, namely photon subtraction and photon addition.
As shown in Fig.~\ref{fig:diagdeng}, both types of operations can be implemented by beam splitters and photon number detectors.
Photon subtraction can be implemented by coupling an input mode with an ancillary vacuum state at a beam splitter with transmissivity approaching one.
The operation is successful if one photon is detected at the ancillary output of the beam splitter.
Let $\hat{a}_1$ and $\hat{a}_2$ be the annihilation operators of the two modes of the TMSV state. 
Then the symmetric photon subtraction can be represented by the operator $\hat{a}_1\hat{a}_2$. 
Photon addition can be implemented in a manner similar to photon subtraction, but with a single-photon state as the ancillary. The operation is successful if no photons are detected at the ancillary of the beam splitter.
The symmetric photon addition can be represented by the operator $\hat{a}^\dagger_1\hat{a}^\dagger_2$.

Beyond the independent applications of symmetric non-Gaussian operations on both modes of the TMSV state, we also consider a delocalized form of non-Gaussian operations, which has been shown to entangle two separable modes \cite{lan2021two}.
As shown in Fig.~\ref{fig:diagdeng}, such operations can also be implemented by beam splitters and photon number detectors.
The major difference is the coherent detection and preparation of the ancillaries.
The delocalized photon subtraction and photon addition can be represented by the operators
$(\hat{a}_1+\hat{a}_2)/\sqrt{2}$ and $(\hat{a}^\dagger_1+\hat{a}^\dagger_2)/\sqrt{2}$, respectively.
When truncated to the space spanned by $\{\ket{0}, \ket{1}\}^2$, the TMSV resource state after the delocalized photon subtraction or photon addition approaches the Bell state $\ket{\Psi^+}$.
A different correction strategy will be adopted in the H-BSM protocol for such a non-Gaussian resource state.
Different from their symmetric counterparts, both delocalized operations can only be applied to the TMSV state before channel transmission because of the coherent detection on the ancillaries.

\begin{figure}
	\centering
	\includegraphics[width=.9\linewidth]{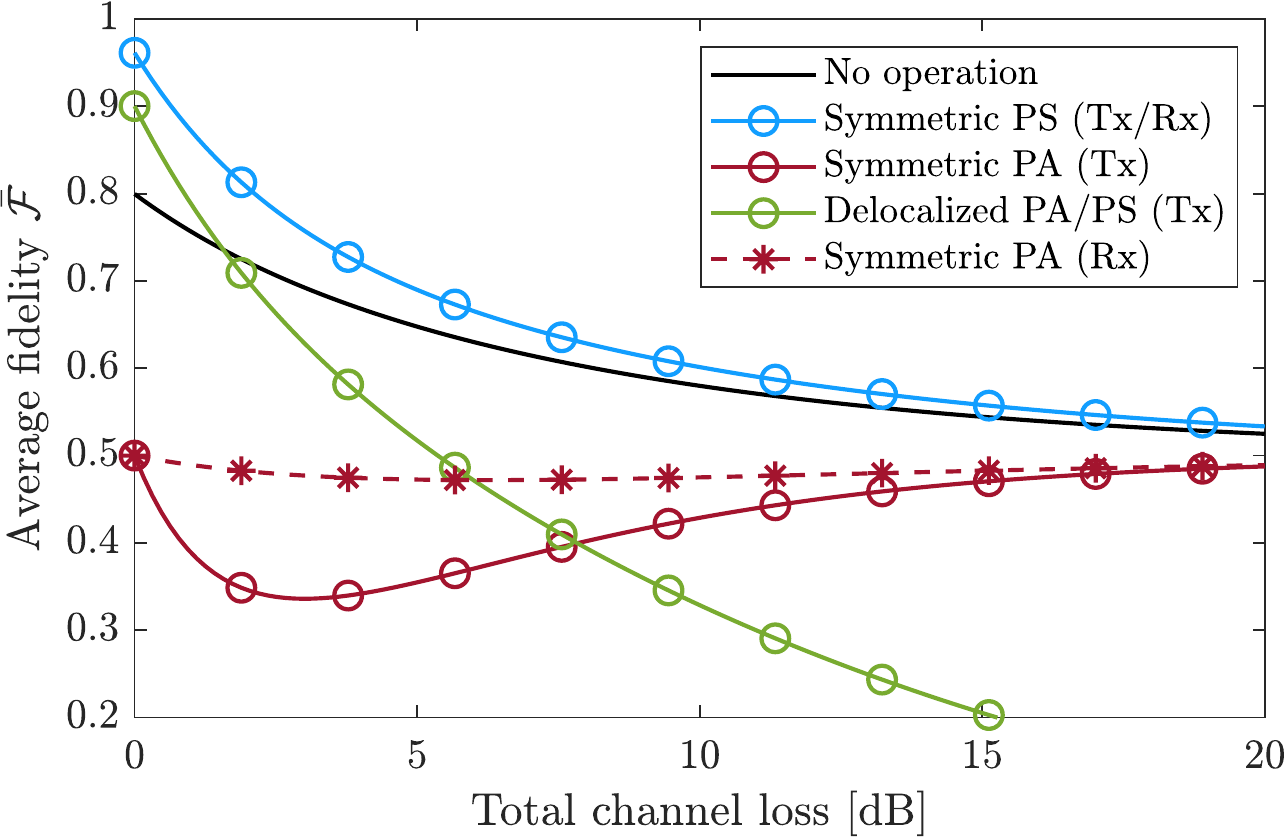}\\
		(a) Initial squeezing $=3$dB\\
		\includegraphics[width=.9\linewidth]{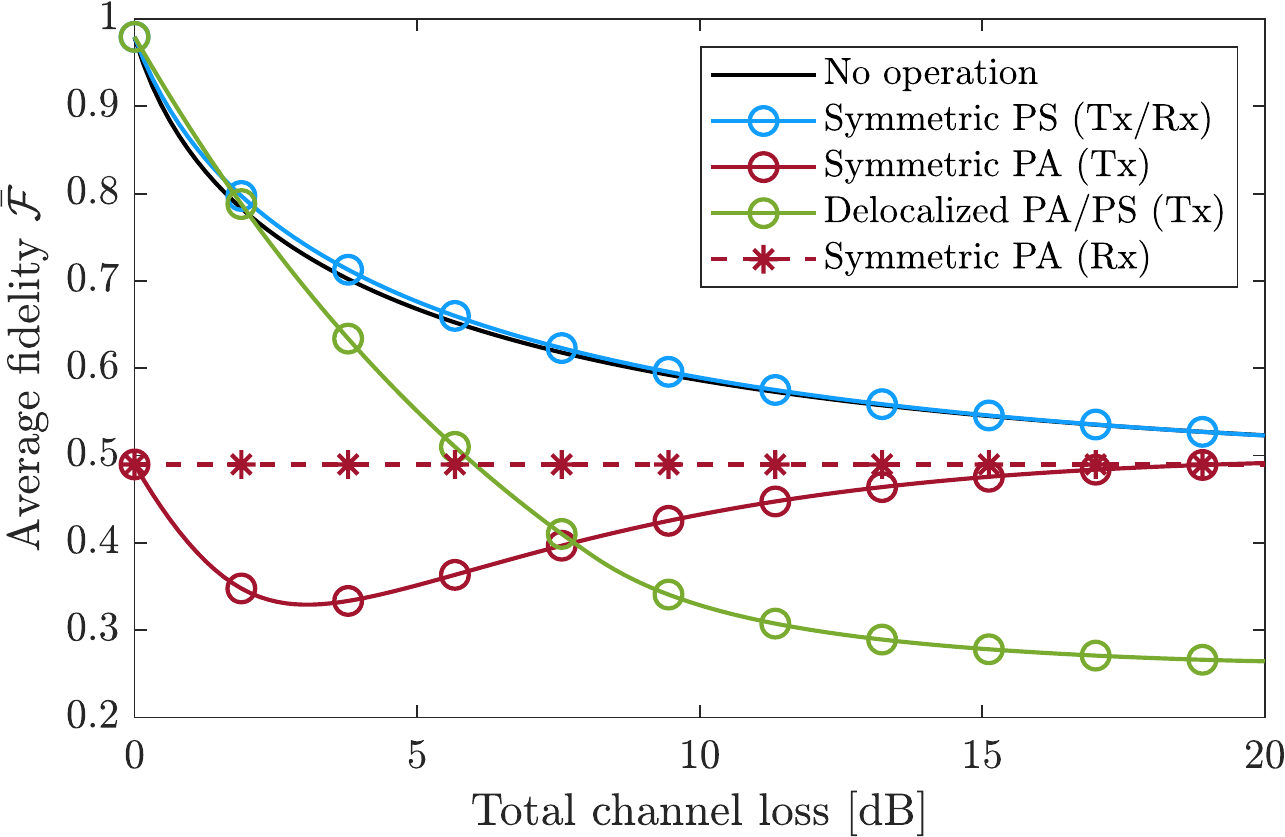}\\
		(b) Initial squeezing is optimized
	\caption{The average fidelity for the resultant hybrid entangled state, of which the CV qubit is teleported using the H-BSM protocol and non-Gaussian operations. 
		The squeezing of the initial TMSV state is fixed in (a), and set to maximize $\bar{\mathcal{F}}$ in (b).
	The non-Gaussian operations are applied to the teleportation resource state before (curves marked by circles and labeled Tx) or after (curves marked by stars and labeled Rx) the state is transmitted over the lossy channels. 
	For the hybrid entangled state $\alpha=0.5$ is set (PS: photon subtraction, PA: photon addition).}
	\label{fig:figngswapping}
\end{figure}

Fig.~\ref{fig:figngswapping} compares the average fidelity for the H-BSM protocol with photon subtraction or photon addition being applied on the resource state.
In Fig.~\ref{fig:figngswapping}a the squeezing of the initial TMSV states is set to a fixed value.
Performing standard (i.e., not delocalized) symmetric photon subtraction on a TMSV state before the channel transmission produce the same fidelity as performing the same operation after the transmission.
This is because the annihilation operator commutes with the operator given by Eq.~(\ref{eq:channelkraus}) (up to some normalization factor).
Performing delocalized photon subtraction produce the same fidelity as delocalized photon addition due to the symmetry of the initial TMSV state.
The delocalized operations provide the largest $\bar{\mathcal{F}}$ when both the initial squeezing and the channel loss is small.
In Fig.~\ref{fig:figngswapping}b the squeezing of the initial TMSV states are set independently so as to maximize $\bar{\mathcal{F}}$ for each operations.
For the non-Gaussian operations considered, only symmetric photon subtraction, delocalized photon addition, and delocalized photon subtraction can improve the fidelity relative to not performing any operation.
However, such operations only provide a few percent improvements with optimized initial squeezing.

Next, consider two other types of non-Gaussian operations, namely photon catalysis and quantum scissors.
Both  operations can be used for the amplification of the single photon component of a state \cite{ralph2009nondeterministic,ulanov2015undoing,zhang2018photon,hu2019entanglement,he2021noiseless}. 
As shown in Fig.~\ref{fig:diagdeng},
photon catalysis can be implemented by using a single photon ancillary state and detecting a  single photon state at the ancillary output.
Let $T_\mathrm{c}$ be the transmissivity of the beam splitter in photon catalysis, then the symmetric photon catalysis can be represented by $\hat{R}_1\hat{R}_2$, where \cite{hu2016multiphoton}
\begin{equation}
	\hat{R}_k=\sqrt{T_\mathrm{c}}\left(\frac{T_\mathrm{c}-1}{T_\mathrm{c}}\hat{a}_k^\dagger\hat{a}_k+1\right)\sqrt{T_\mathrm{c}}^{\hat{a}_k^\dagger\hat{a}_k},\ k=1,2.
\end{equation}
The two terms in the parenthesis are due to the indistinguishability of the source of the detected single photon.
Different from photon subtraction and photon addition, $\hat{R}_k$ approaches the identity operator when $T_\mathrm{c}$ approaches one.

\begin{figure}
	\centering
	\includegraphics[width=0.45\linewidth]{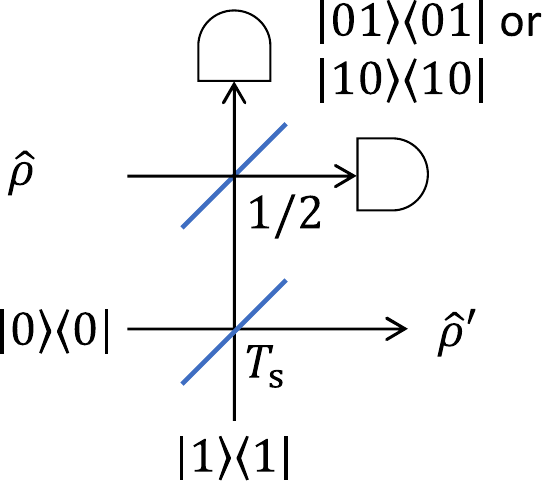}
	\caption{The quantum scissors is applied to one mode of an input $\hat{\rho}$. The operation has been successful if one photon is detected at either output of the 50:50 beam splitter. On the detection of $\ket{10}$, a phase shift will be applied to the output mode.}
	\label{fig:diagqs}
\end{figure}

As shown in Fig.~\ref{fig:diagqs}, quantum scissors consists of a DV entangled state and an incomplete H-BSM.
The entangled state is created by mixing a single photon state and a vacuum state at a beam splitter with transmissivity $T_\mathrm{s}$.
The incomplete H-BSM is implemented by a 50:50 beam splitter and two photon number detectors.
The symmetric quantum scissors can be represented by $\hat{M}_1\hat{M}_2$, where \cite{ralph2009nondeterministic}
\begin{equation}
		\hat{M}_k=\sqrt{T_\mathrm{s}}\ket{0}_k\bra{0}+\sqrt{1-T_\mathrm{s}}\ket{1}_k\bra{1},\ k=1,2.
\end{equation}

\begin{figure}
	\centering
	\includegraphics[width=.9\linewidth]{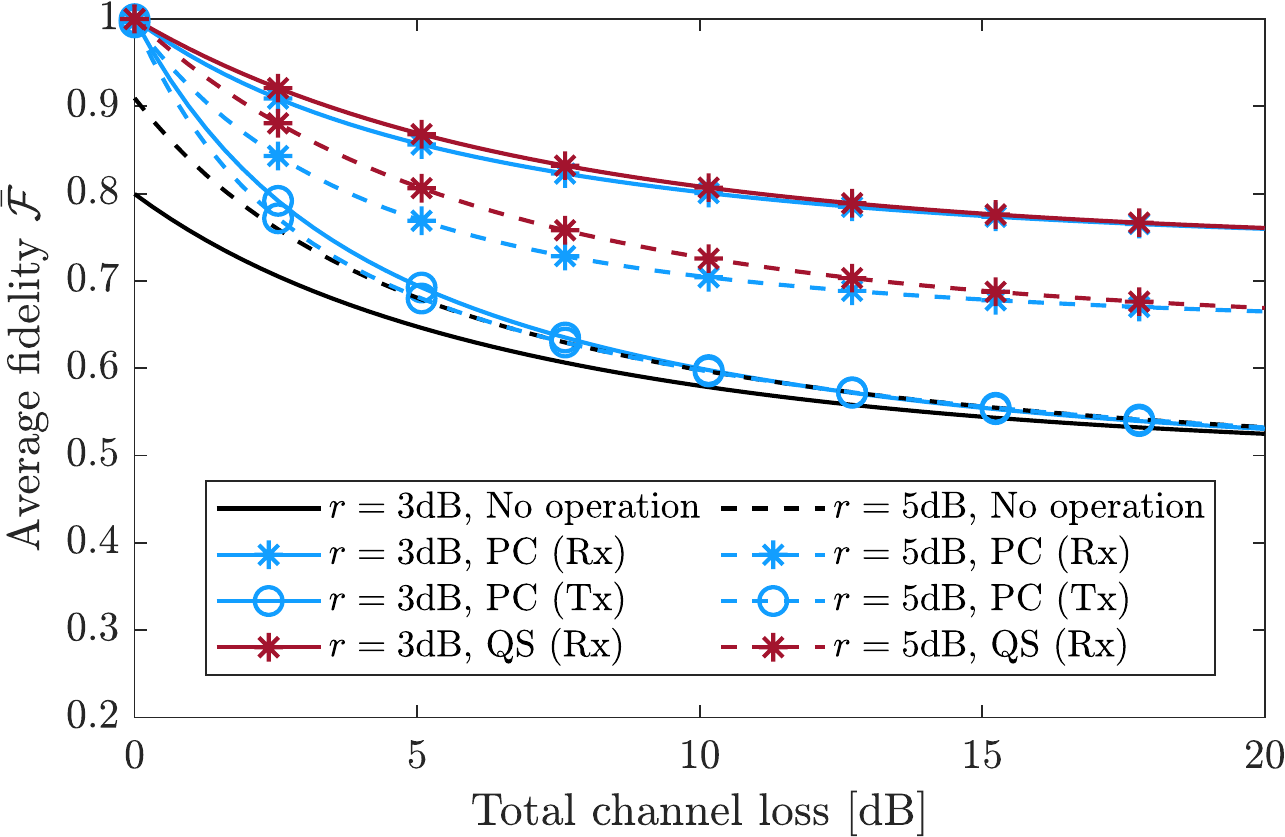}
	\caption{The maximized average fidelity for the resultant hybrid entangled state, of which the CV qubit is teleported using the H-BSM protocol and symmetric non-Gaussian operations.
	For the hybrid entangled state $\alpha=0.5$ is set (PC: photon catalysis, QS: quantum scissors).}
	\label{fig:figbellandhomoqsandpc}
\end{figure}

Fig.~\ref{fig:figbellandhomoqsandpc} compares the average fidelity for the H-BSM protocol with symmetric photon catalysis and quantum scissors.
Quantum scissors are applied only after the TMSV states are distributed over symmetric lossy channels.
For both operations the beam splitter transmissivities $T_\mathrm{c}$ and $T_\mathrm{s}$ are set independently so as to maximize $\bar{\mathcal{F}}$.
We find that both operations can improve the average fidelity of the protocol when the initial squeezing of the TMSV state is small.
For a given channel loss and an initial squeezing, quantum scissors provide higher fidelity than photon catalysis.
Performing photon catalysis at the receiver always provides fidelity higher than the sender.
Although not shown, for quantum scissors, which provides the highest fidelity, the total success probability for the H-BSM protocol is $\sim10^{-2}$ when the channel loss is above 10dB.

\section{Conclusion}\label{sec:5}
In this work, we investigated the feasibility of teleporting different quantum states using a new teleportation protocol, the H-BSM protocol.
We first found that, relative to the traditional teleportation protocol based on CV-BSM, teleportation of CV-only qubits states can be enhanced under our new protocol. The enhancement decreases as the mean photon number of the CV qubit state grows and vanishes as that number exceeds a certain threshold. This conclusion is true for any fixed channel loss, but the threshold is found to depend on the loss value.

We then turned to our main focus---the teleportation of a hybrid entangled state.
We compared our protocol with the traditional protocol using the achievable fidelity under fixed  channel loss. 
In teleporting the DV qubit of the hybrid entangled state, we found the H-BSM protocol always provides higher fidelity than the CV-BSM protocol.
In teleporting the CV qubit of the hybrid entangled state, we found that no protocol is always superior. The H-BSM protocol outperforms the CV-BSM protocol only when the mean photon number of the CV qubit is below a certain threshold. 
For both protocols, we found teleporting the DV qubit of the hybrid entangled state using either protocol can always achieve higher fidelity than teleporting the CV qubit of the state.

We also investigated the use of different non-Gaussian operations in the H-BSM protocol as applied to our hybrid entangled state.
The operations we studied included photon subtraction, photon addition, photon catalysis, and quantum scissors. 
We considered different scenarios where the operations were applied (in both modes) before or after  the distribution of the TMSV resource states. We also considered scenarios where the subtraction and addition at each mode was delocalized.
For hybrid entangled states with small mean photon numbers in their CV qubits, we found that all operations considered improved fidelity under certain conditions, at the cost of reduced success probability. Quantum scissors always provided the most improvement.
\bibliographystyle{mybst}
\bibliography{mybib}

\end{document}